\documentclass[prx,aps,superscriptaddress,floatfix,tightenlines,reprint]{revtex4-2}
\usepackage{amsthm}
\usepackage{amsmath}
\usepackage{amssymb}
\usepackage{amsfonts}
\usepackage{epsfig}
\usepackage{bm}
\usepackage{times}
\usepackage[colorlinks,linkcolor=blue,anchorcolor=blue,citecolor=blue,urlcolor=blue]{hyperref}
\usepackage{braket}
\usepackage{graphicx}
\usepackage{setspace}
\usepackage{subfigure}
\usepackage{enumerate}

\theoremstyle{definition}

\newtheorem*{definition*}{Definition}

\theoremstyle{plain}

\newtheorem*{theorem*}{Theorem}
\newtheorem*{corollary*}{Corollary}
\newtheorem*{lemma*}{Lemma}
\newtheorem*{proposition*}{Proposition}

\NewDocumentEnvironment{variant}{O{theorem} D(){} m}
    {\addtocounter{#1}{-1}%
    \expandafter\renewcommand\csname the#1\endcsname{\ref{#3}$'$}%
    \begin{#1}[#2]}
    {\end{#1}}

\NewDocumentEnvironment{appthm}{O{theorem} D(){} m}
    {\addtocounter{#1}{-1}%
    \expandafter\renewcommand\csname the#1\endcsname{\ref{#3}}%
    \begin{#1}[#2]}
    {\end{#1}}

\begin{document}

    \title{Distributed Quantum Computation via Entanglement Forging and Teleportation}

    \author{Tian-Ren Jin}
        \affiliation{Institute of Physics, Chinese Academy of Sciences, Beijing 100190, China}
        \affiliation{School of Physical Sciences, University of Chinese Academy of Sciences, Beijing 100049, China}

    \author{Kai Xu}
        \affiliation{Institute of Physics, Chinese Academy of Sciences, Beijing 100190, China}
        \affiliation{School of Physical Sciences, University of Chinese Academy of Sciences, Beijing 100049, China}
        \affiliation{Beijing Academy of Quantum Information Sciences, Beijing 100193, China}
        \affiliation{Hefei National Laboratory, Hefei 230088, China}
        \affiliation{Songshan Lake Materials Laboratory, Dongguan 523808, China}
        \affiliation{CAS Center for Excellence in Topological Quantum Computation, UCAS, Beijing 100190, China}

    \author{Heng Fan}
        \email{hfan@iphy.ac.cn}
        \affiliation{Institute of Physics, Chinese Academy of Sciences, Beijing 100190, China}
        \affiliation{School of Physical Sciences, University of Chinese Academy of Sciences, Beijing 100049, China}
        \affiliation{Beijing Academy of Quantum Information Sciences, Beijing 100193, China}
        \affiliation{Hefei National Laboratory, Hefei 230088, China}
        \affiliation{Songshan Lake Materials Laboratory, Dongguan 523808, China}
        \affiliation{CAS Center for Excellence in Topological Quantum Computation, UCAS, Beijing 100190, China}

    \begin{abstract}
        Distributed quantum computation is a practical method for large-scale quantum computation on quantum processors with limited size.
        It can be realized by direct quantum channels in flying qubits.
        Moreover, the pre-established quantum entanglements can also play the role of quantum channels with local operations and classical channels.
        However, without quantum correlations like quantum channels and entanglements, the entanglement forging technique allows us to classically forge the entangled states with local operations and classical channels only.
        In this paper, we demonstrate the methods to implement a nonlocal quantum circuit on two quantum processors without any quantum correlations, which is based on the fact that teleportation with classically forged Bell states is equivalent to quantum state tomography.
        In compensation, the overhead of single-shot measurement will increase, and several auxiliary qubits are required.
        Our results extend the possibility of integrating quantum processors.
        We expect that our methods will complement the toolbox of distributed quantum computation, and facilitate the extension of the scale of quantum computations. 
    \end{abstract}
    \maketitle

    \section{Introduction}

    Quantum computation has attracted widespread attention in recent years, since it takes advantage when simulating large quantum systems, or solving specific problems with efficient quantum algorithms.
    Practical quantum computation requires millions of qubits in a relatively low level of noise, which may obstruct the applications of quantum computation.
    Quantum error correction and other techniques provide a systematic way to deal with the noises in quantum operations~\cite{RevModPhys.87.307,RevModPhys.95.045005}.
    Recently, the remarkable progress in quantum error correction codes has declared that the level of noise in the state-of-art experimental techniques is close to the threshold of fault-tolerant quantum computation~\cite{PhysRevLett.129.030501,google2023suppressing,evered2023high}. 

    However, fabricating such many qubits on an individual quantum chip is challenging for state-of-the-art experimental techniques.
    Quantum internet and distributed quantum computation provide a feasible way to extend the size of quantum processors~\cite{RevModPhys.95.045006,luo2023recent,fang2023quantum}.
    On many integrated quantum processors, the scales of implementable quantum tasks can be much larger than those on an individual quantum processor.

    To integrate two spatially separate quantum processors, it can be directly correlated with flying qubits to transmit quantum states, which have been realized by both photons~\cite{kurpiers2018deterministic,PhysRevLett.120.200501,zhong2019violating,zhong2021deterministic,PhysRevLett.132.047001} and phonons~\cite{bienfait2019phonon,zivari2022chip}.
    The famous quantum teleportation protocol also shows that quantum channels can be undertaken by entangled states with classical channels~\cite{PhysRevLett.70.1895}. 
    In addition, the gate teleportation protocol teleports the nonlocal controlled-unitary quantum gates with the assistance of only a single Bell pair~\cite{PhysRevA.62.052317,hu2023progress}.
    Therefore, nonlocal circuits can be implemented on separate quantum processors with quantum correlations.

    However, when two quantum processors are separated with only local operations and classical channels (LOCCs), the nonlocal quantum circuits can still be implemented.
    Recently, a technique called entanglement forging has been used to double the size of quantum simulators~\cite{PRXQuantum.3.010309}.
    In this technique, entangled states can be forged by separable states classically.
    Entanglement forging has been used only to double the size of quantum processors in some specially designed circuits~\cite{PhysRevResearch.6.023021,huembeli2022entanglement}.
    With the entanglement forging and the teleportation protocol, any nonlocal quantum circuits can be forged on many separate quantum processors with only classical channels.
    
    This paper is organized as follows. 
    In Sec.~\ref{sec: prelimilaries}, we review the entanglement forging and the gate teleportation protocol.
    In Sec.~\ref{sec: distributed}, we show that the quantum state teleportation with entanglement forging is equivalent to quantum state tomography, and demonstrate the methods to forge a nonlocal quantum circuit on two separated quantum processors with LOCCs. 
    In Sec.~\ref{sec: noise}, we discuss the suppression of measurement noise in our methods with measurement error mitigation.
    The conclusion and discussion are given in Sec.~\ref{sec: conclusion}.

    \section{Prelimilaries} \label{sec: prelimilaries}

    \subsection{Entanglement Forging}

        Entanglement forging employs the Schmidt decomposition of a bipartite entangled state
        \begin{equation}
            \ket{\psi} = (\hat{U} \otimes \hat{V}) \sum_{i} \lambda_i \ket{i} \otimes  \ket{i},
        \end{equation} 
        where $\ket{i}$ is the computational basis of local Hilbert space, and $\hat{U}$ and $\hat{V}$ are unitaries, and the Schmidt coefficients $\lambda_i$ are positive.
        In terms of the density matrix, it can be written as
        \begin{align}
            \rho_{\psi} & = (\hat{U} \otimes \hat{V}) \sum_i \left(\lambda_i^2 \ket{i}\bra{i} \otimes \ket{i}\bra{i} + \sum_{j<i} \right. \nonumber\\
            &~~~~\left.\times \sum_{p \in \mathbb{Z}_4} (-1)^p \ket{ij_p} \bra{ij_p} \otimes \ket{ij} \bra{ij_p} \right) (\hat{U}^{\dagger} \otimes \hat{V}^{\dagger}),
        \end{align} 
        where $\ket{ij_p} = \frac{1}{\sqrt{2}}(\ket{i} + \mathrm{i}^p \ket{j})$, with $p \in \{1,2,3,4\}$.

        In this decomposition, the entangled state $\ket{\psi}$ is not the classical probabilistic mixture of separable states, where all the coefficients of the decomposition are positive, due to the factor $(-1)^p$.
        This kind of decomposition is called local pesudomixture~\cite{PhysRevA.58.826,PhysRevA.59.141} in the investigation of the robustness measure of quantum entanglement.
        In general, we can decompose a state $\rho_{AB}$ of system $AB$ into the product states of $AB$
        \begin{equation} \label{eq: decomposition}
            \rho_{AB} = \sum_i x_i \rho_A^i \otimes \rho_B^i,
        \end{equation} 
        where the coefficients $x_i$ can be both positive and negative.
        With the normalization of states, the coefficients satisfy $\sum_i x_i = 1$, thus this decomposition is called the quasiprobability decomposition. 

        If all the coefficients are positive, this state is a separable state by definition.
        Therefore, for an entangled state, there exist coefficients $x_i <0$.
        The decomposition of the entangled state can be rewritten as 
        \begin{equation}
            \rho_{AB} = Z \sum_i \mathrm{sgn}(x_i) q_i \rho_A^i \otimes \rho_B^i,
        \end{equation}
        where $Z = \sum_i |x_i|$, and $q_i = |x_i|/Z$ is a probabilistic distribution.
        Therefore, the expectation of observable $\hat{O}$ in $\rho_{AB}$ is
        \begin{equation}
            \braket{\hat{O}}_{\rho_{AB}} = Z \sum_i q_i \left(\mathrm{sgn}(x_i) \braket{\hat{O}}_{\rho_A^i \otimes \rho_B^i}\right),
        \end{equation}
        the probabilistic mixture of the expectations in separable state $\rho_A^i \otimes \rho_B^i$ with signature $\mathrm{sgn}(x_i)$.
        The entangled state can be simulated by a separable state in the sense of expectations.
        A similar technique is also used in the probabilistic error cancellation method of quantum error mitigation~\cite{RevModPhys.95.045005,PhysRevLett.119.180509,PhysRevX.8.031027,cai2021multi,PhysRevResearch.3.033178,van2023probabilistic}.

        The cost of separable states in the simulation of an entangled state is evaluated as the factor $Z$, and the minimal cost over all possible decomposition is called the implementability of the state $\rho_{AB}$ with respect to separable states~\cite{jin2024noisy}
        \begin{equation}
            p_{\mathcal{Q}}(\rho_{AB}) = \min \left\{\sum_i |x_i|| \rho_{AB} = \sum_i x_i \rho_A^i \otimes \rho_B^i\right\}.
        \end{equation}

    \subsection{Gate Teleportation}

        \begin{figure*}[t]
            \centering
            \includegraphics{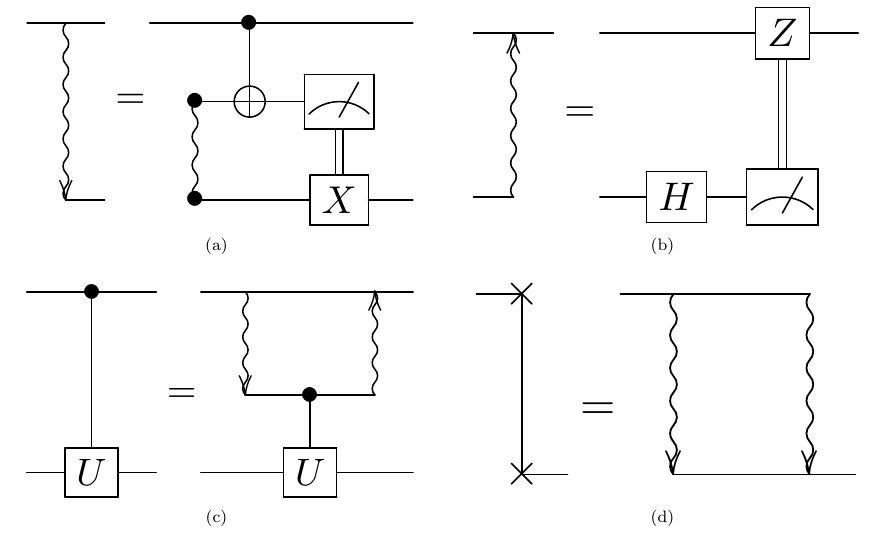}
            \caption{The circuits of (a) the starting process, (b) the ending process, (c) the gate teleportation of the controlled-unitary gate, and (d) the state teleportation.
            The wavy line connecting two qubits in (a) represents a Bell pair.
            }
            \label{fig: teleportation}
        \end{figure*}

        Gate teleportation protocol is put forward to nonlocally implement the quantum CNOT gate with the assistance of a pair of qubits in the maximal entangled state.
        It can also be extended to teleport any $N$-controlled unitary gate.
        Given a controlled-unitary gate $\mathcal{U}_C$, there are universal starting and ending processes~\cite{wu2023entanglement}, which allows for nonlocal application of the gate $\mathcal{U}_C$ with the consumption of only a single Bell pair.
        On the contrary, for the nonlocal implementation of a general two-bit gate, with the teleportation of state, it consumes two Bell pairs. 

        The gate teleportation protocol of a controlled-unitary gate $\mathcal{U}_C$ is shown in Fig.~\ref{fig: teleportation}(c).
        The starting and ending process is part of the teleportation of state, as shown in Fig.~\ref{fig: teleportation}(d), which contains the Bell measurement and the feedback.   
        Assume the input state of the starting process is $\ket{\psi_{\mathrm{in}}} = \alpha \ket{0} + \beta \ket{1}$, which is the state of a qubit called the control qubit in the following. 
        Then the output state of the starting process is  
        \begin{equation}
            \ket{\psi_{\mathrm{start}}} = \alpha \ket{00} + \beta \ket{11},
        \end{equation}
        where the second qubit is an auxiliary qubit.
        The ending process can take the state $\ket{\psi_{\mathrm{start}}}$ back to $\ket{\psi_{\mathrm{in}}}$, without the cost of nonlocal quantum operations or quantum state.
        
        If a two-qubit quantum gate $\hat{U}$ is acted before the ending process, the action of the output state will generally not be the two-qubit quantum gate $\hat{U}$ 
        Expand the quantum gate $\hat{U}$ into the Pauli basis on the auxiliary qubit after starting the process as
        \begin{equation}
            \hat{U} = \sum_{i=0,x,y,z} \hat{\sigma}_i \otimes \hat{U}_i.
        \end{equation}
        The action of this gate is 
        \begin{align}
            & \hat{I} \otimes \hat{U} \ket{\psi_{\mathrm{start}}} \otimes \ket{\psi_{\mathrm{t}}} \nonumber \\
            & = \sum_i (\alpha \ket{0}\otimes \sigma_i \ket{0} + \beta \ket{1}\otimes \sigma_i \ket{1})\otimes \hat{U}_i \ket{\psi_{\mathrm{t}}},
        \end{align}    
        where $\ket{\psi_{\mathrm{t}}}$ is the state of qubit not involved in the starting and ending process, which is called target qubit in the following.
        After the ending process of the measurement in $x$ direction on auxiliary qubit and feedback, the state on control and target qubit is 
        \begin{align}
            \ket{\psi_{\mathrm{out}}^+} & = [\hat{I} \otimes(\hat{U}_0 + \hat{U}_x) + \hat{\sigma}_z \otimes(\hat{U}_z + \mathrm{i} \hat{U}_y )]\ket{\Psi_{\mathrm{in}}} , \\
            \ket{\psi_{\mathrm{out}}^-} & = [\hat{I} \otimes(\hat{U}_0 - \hat{U}_x) + \hat{\sigma}_z \otimes(\hat{U}_z -\mathrm{i} \hat{U}_y)]\ket{\Psi_{\mathrm{in}}} ,
        \end{align}
        where $\ket{\Psi_{\mathrm{in}}} = \ket{\psi_{\mathrm{in}}} \otimes \ket{\psi_{\mathrm{t}}}$.
        To realize the gate teleportation, it requires that
        \begin{equation}
            \ket{\psi_{\mathrm{out}}^+} = \ket{\psi_{\mathrm{out}}^-} = \hat{U} \ket{\Psi_{\mathrm{in}}},
        \end{equation}
        which gives that 
        \begin{equation}
            \hat{U}_x = \hat{U}_y = 0.
        \end{equation}
        This means that the quantum gates, which can be teleported with the starting and ending process, are in the form
        \begin{equation}
            \hat{U} = \hat{I}\otimes \hat{U}_0 + \hat{U}_{\mathrm{c}} \hat{\sigma}_z \hat{U}_{\mathrm{c}}^{\dagger} \otimes \hat{U}_z,
        \end{equation}
        where $\hat{U}_{\mathrm{c}}$ is arbitrary unitary transformation acting on the control qubit.
        In this protocol, the gate teleportation costs one pair of qubits in the Bell state.
        For the general two-qubit quantum gate, the gate teleportation is realized with the state teleportation twice, which costs two pairs of Bell state.  

    \section{Distributed Quantum Computation with Local Operations and Classical Channels} \label{sec: distributed}

    \subsection{Teleportation Forged by Separable States}
    
    The teleportation protocol teleports qubits with classical channels in the consumption of Bell state~\cite{PhysRevLett.70.1895}, so the entangled states are resources for quantum communication. 
    The entangled states cannot be prepared by local operations and classical channels (LOCC), which is the free set of the quantum entanglement theory~\cite{PhysRevA.53.2046,PhysRevA.54.3824,PhysRevLett.78.2275}.
    However, we can simulate the entangled states $\rho^{E}$ by the quasiprobability decomposition.

    Here is an example of the simulation of the Bell state.
    \begin{align} 
        \ket{B^+} \bra{B^+} =& \frac{1}{2} \left(\ket{00} \bra{00} + \ket{00} \bra{11} \right. \nonumber\\
        & + \left. \ket{11} \bra{00} + \ket{11} \bra{11}\right) \nonumber\\
        =& \frac{1}{4} \left(\hat{I} + \hat{Z}_1 \hat{Z}_2 + \hat{X}_1 \hat{X}_2 - \hat{Y}_1 \hat{Y}_2 \right) \nonumber\\
        =& \frac{1}{4} \left(\hat{I} + \hat{Z}_1 \hat{Z}_2\right) + \frac{1}{4} \left(\hat{I} + \hat{X}_1 \hat{X}_2\right) \nonumber\\
        & - \frac{1}{4} \left(\hat{I} + \hat{Y}_1 \hat{Y}_2\right), \label{eq: bell_state}
    \end{align} 
    where
    \begin{align*}
        \frac{1}{4} \left(\hat{I} + \hat{Z}_1 \hat{Z}_2\right) = &\frac{1}{2} \left(\ket{0_z0_z} \bra{0_z0_z} + \ket{1_z1_z} \bra{1_z1_z}\right) ,\\
        \frac{1}{4} \left(\hat{I} + \hat{X}_1 \hat{X}_2\right) = &\frac{1}{2} \left(\ket{0_x0_x} \bra{0_x0_x} + \ket{1_x1_x} \bra{1_x1_x}\right) ,\\
        \frac{1}{4} \left(\hat{I} + \hat{Y}_1 \hat{Y}_2\right) = & \frac{1}{2} \left(\ket{0_y0_y} \bra{0_y0_y} + \ket{1_y1_y} \bra{1_y1_y}\right) .
    \end{align*}

    The teleportation protocol is implemented by appending a state $\rho_C$ and performing the Bell state measurements $\hat{\Pi}_{s}, s=0,1,2,3$ on the system $CA$.
    If the measurement outcome is $s_0$, then the state is 
    \begin{align}
        \hat{\Pi}_{s_0} \rho_C \otimes \rho_{AB} \hat{\Pi}_{s_0} & = \sum_i x_i \hat{\Pi}_{s_0} \rho_C \otimes \rho_{Ai} \otimes \rho_{Bi} \hat{\Pi}_{s_0} \\
        & = \hat{\Pi}_{s_0} \otimes \sum_i x_i p(s_0|i) \rho_{Bi},
    \end{align}
    where $p(s|i) = \mathrm{Tr}\left(\hat{\Pi}_{s} \rho_C \otimes \rho_{Ai}\right) = \mathrm{Tr}\left(\rho_{Ai}^{T} \sigma_{s} \rho_C \sigma_{s}\right)$.
    The transported state $\rho_C$ is obtained by 
    \begin{equation}
        \rho_C \propto \sum_i x_i p(s_0|i) \mathcal{U}_{s_0}(\rho_{Bi}) = \sum_s \sum_i x_i p(s|i) \mathcal{U}_{s}(\rho_{Bi}).
    \end{equation}
    Thus, with decomposition Eq.~(\ref{eq: decomposition}), the classical information of $p(s|i)$ for a fixed $s$ is sufficient to recover the state $\rho_C$.
    Therefore, the teleportation protocol can be simulated by separable states completely, which means that the quantum information of Alice is converted to Bob completely by classical channels.

    For simplicity, we consider $\sigma_{s=0} = I$, then $p(0|i) = \mathrm{Tr}\left(\rho_{Ai}^{T} \rho_C \right)$.
    In the decomposition Eq.~(\ref{eq: bell_state}), $\rho_{Ai} = \ket{i_{\alpha}} \bra{i_{\alpha}}$, where $i = 0,1$ and $\alpha = x, y, z$, so as $\rho_{Ai}^{T}$.
    Therefore, the information of $p(0|i)$ is the same as the required information of standard quantum state tomography (QST) on single qubit~\cite{nielsen2010quantum}, so the teleportation with the Bell state simulated in this decomposition is equivalent to the standard QST. 
    Because the dimension of the single qubit state space is $d = 3$, which is equal to the number of quantities measured by standard QST, it is reasonable to believe that the implementability is attained by this decomposition, $p_{\mathcal{S}}(\rho_{AB}) = 3$.

    When simulating many Bell pairs, let the system of Bell states $\rho_{AB}$ be $AB = \bigotimes_{a} A_{a} B_{a}$, where $A_{a}, B_{a}$ are single qubit.
    If the state that can be freely used is the separable state in the systems $A_a$ and $B_a$,  $\mathcal{F}_{\mathcal{S}} = \mathrm{conv}\left[\bigotimes_{a} \mathcal{S}(A_{a}B_{a})\right]$, it can be shown that (Proposition~4 in Ref.~\cite{jin2024noisy})
    \begin{equation} \label{eq: local_tomography}
        p_{\mathcal{F}_{\mathcal{S}}}(\rho_{AB}) = \prod_a p_{\mathcal{S}}(\rho_{A_a B_a}) = 3^{L},
    \end{equation}
    where $L$ is the number of simulated Bell pairs.
    The teleportation with Bell pairs in this simulation is equivalent to the local tomography~\cite{barnum2014local,chiribella2016quantum}.
    Equation~(\ref{eq: local_tomography}) implies that the standard QST is the most efficient scheme in local tomography.

    In the contrary, if the entangled states between system $A_a$ (or system $B_a$) can be freely used, $\mathcal{F} = \mathcal{S}(AB) = \mathrm{conv}\left[\mathcal{Q}(A) \otimes \mathcal{Q}(B)\right]$, it can be show that (Proposition~5 and Proposition~11 in Ref.~\cite{jin2024noisy})
    \begin{equation}
        2^L \leq p_{\mathcal{S}}(\rho_{AB}) \leq 3^{L}.
    \end{equation}
    The investigations in classical teleportation also show that the standard QST is not the most efficient scheme in global tomography~\cite{PhysRevLett.74.1259,hu2023progress}.

    With the classically forged Bell states and the teleportation protocols, a quantum computational task can be performed on separate quantum processors with only classical communications.
    In compensation, there are exponentially increasing overhead of repeated measurements.
    For an arbitrary quantum gate, its teleportation requires two pairs of qubits in the Bell state.
    In the decomposition of Eq.~(\ref{eq: bell_state}), the overhead for classical teleportation of one nonlocal quantum gate is $9$~times more than the performing on one chip, and $4$~auxiliary qubit.
    However, to teleport a general two-qubit gate, it requires two pairs of qubits in Bell states.
    In this case, the entanglement forging in Eq.~(\ref{eq: bell_state}) is not optimal, and the optimal one has overhead larger than $4$~times.
    Since the classical teleportations of nonlocal quantum gates are independent, assume that the quantum circuit has $L$~layers, where each layer contains one nonlocal gate between two processors, the $p^{2L}$~times overhead and the $4 L$~auxiliary qubits are required, where $2 \leq p \leq 3$.

    \subsection{Identical operations Forged by Projective Measurements}

    \begin{figure*}[t]
        \centering
        \includegraphics{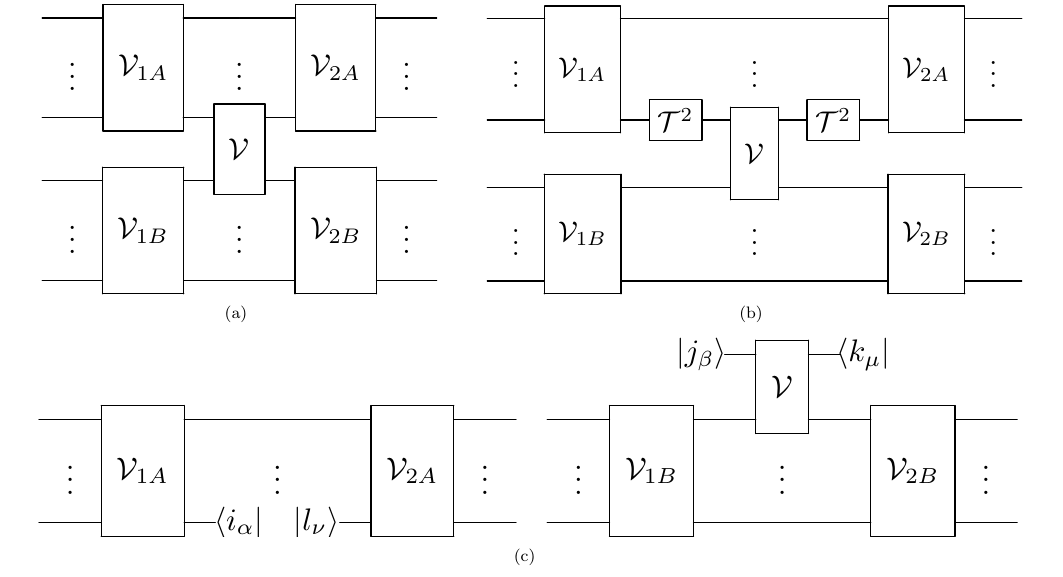}
        \caption{(a) The nonlocal circuit. (b) Separating the circuit by inserting $\mathcal{I} = \mathcal{T}^2$. (c) Separated circuits of Alice and Bob. The separate circuits in (c) are paired with the transition matrix $M_{j\beta,i\alpha}$ and $M_{l\nu,k\mu}$.
        }
        \label{fig: division}
    \end{figure*}

        Since the distributed quantum computation is aimed at scaling the size of the quantum computer, the increasing auxiliary qubit will confine its usage.
        In the following, we illustrate a method with less number of auxiliary qubits.

        The state teleportation can be viewed as an identical operation between two qubits.
        In the state teleportation with entanglement forging, one auxiliary qubit is needed to realize the identical operations.
        Therefore, we hope to classically simulate this operation directly.

        In terms of the Choi-Jamiołkowsky (CJ) isomorphism~\cite{JAMIOLKOWSKI1972275,CHOI1975285}, the Bell state is dual to the identical operations
        \begin{align} 
            \ket{B^+} \bra{B^+} =& \frac{1}{2} \left(\ket{00} \bra{00} + \ket{00} \bra{11} \right. \nonumber\\
            & + \left. \ket{11} \bra{00} + \ket{11} \bra{11}\right) \nonumber\\
            \mapsto \mathcal{I}(\cdot) = & \ket{0}\!\bra{0}(\cdot) \ket{0}\!\bra{0} + \ket{0}\!\bra{0}(\cdot) \ket{1}\!\bra{1} \nonumber\\
            & + \ket{1}\!\bra{1}(\cdot) \ket{0}\!\bra{0} + \ket{1}\!\bra{1}(\cdot) \ket{1}\!\bra{1}.
        \end{align} 
        This inspires us to construct the identical operations from the classical forging of Bell state, Eq.~(\ref{eq: bell_state}), by transposing the ``bra'' and ``ket'' in the middle of the states.
        This construction gives 
        \begin{align}
            \mathcal{I}(\cdot) & =  \hat{\Pi}_{0z}(\cdot)\hat{\Pi}_{0z} + \hat{\Pi}_{1z}(\cdot)\hat{\Pi}_{1z} + \hat{\Pi}_{0x}(\cdot)\hat{\Pi}_{0x} \nonumber \\
            &~~~~ + \hat{\Pi}_{1x}(\cdot)\hat{\Pi}_{1x}- \hat{\Pi}_{01y}(\cdot)\hat{\Pi}_{10y} + \hat{\Pi}_{10y}(\cdot)\hat{\Pi}_{01y} ,
        \end{align}
        where $\hat{\Pi}_{ij\alpha} = \ket{i_{\alpha}}\!\bra{j_{\alpha}}$ and $\hat{\Pi}_{i\alpha} \equiv \hat{\Pi}_{ii\alpha}$.
        It is the mixture of the projective measurements in $x$ and $z$ directions with the operations $[\hat{\Pi}_{01y}(\cdot)\hat{\Pi}_{10y} + \hat{\Pi}_{10y}(\cdot)\hat{\Pi}_{01y}]$ removed.
        The removed operation is not the projective measurement in $y$ direction, since 
        \begin{equation}
            \ket{0_y}^{T} = \bra{1_y}, \quad \ket{1_y}^{T} = \bra{0_y}.
        \end{equation}

        Although, the operations $[\hat{\Pi}_{01y}(\cdot)\hat{\Pi}_{10y} + \hat{\Pi}_{10y}(\cdot)\hat{\Pi}_{01y}]$ is not a projective measurement.
        It is equivalent to projective measurement up to a local unitary transformation
        \begin{equation}
            \hat{\Pi}_{01y}(\cdot)\hat{\Pi}_{10y} + \hat{\Pi}_{10y}(\cdot)\hat{\Pi}_{01y} = \mathcal{P}_z \circ M_{y}(\cdot),
        \end{equation}
        where $\mathcal{P}_z$ is the Pauli $z$ operation.
        Therefore, the identical operation is 
        \begin{equation}
            \mathcal{I}(\cdot) = \mathcal{M}_{z}(\cdot) + \mathcal{M}_{x}(\cdot) - \mathcal{P}_z \circ \mathcal{M}_{y}(\cdot),
        \end{equation}
        where
        \begin{equation}
            \mathcal{M}_{\alpha}(\cdot) = \hat{\Pi}_{0\alpha}(\cdot)\hat{\Pi}_{0\alpha} + \hat{\Pi}_{1\alpha}(\cdot)\hat{\Pi}_{1\alpha}
        \end{equation}
        is projective measurement in $x,y$, and $z$ direction.
        This allows us to simulate the identical channels classically.

        Assume Alice ($A$) and Bob ($B$) want to perform a nonlocal circuit $\mathcal{U}$.
        This circuit can be represented as 
        \begin{equation}
            \mathcal{U} = (\mathcal{V}_{2A} \otimes \mathcal{V}_{2B}) \circ \mathcal{V} \circ (\mathcal{V}_{1A} \otimes \mathcal{V}_{1B}),
        \end{equation}
        where $\mathcal{V}$ is a nonlocal gate acting on one qubit $q_a$ of Alice and one qubit $q_b$ of Bob.
        Let the initial state be $\rho_{A} \otimes \rho_{B}$. 
        Typically, it can be selected as $\ket{0}_A^{\otimes m} \otimes \ket{0}_B^{\otimes n}$ in practice.
        The output state is
        \begin{equation}
            \rho_{\mathrm{out}} = \mathcal{U}(\rho_{A} \otimes \rho_{B}).
        \end{equation}

        Insert $\mathcal{I}$ before and after the gate $\mathcal{V}$ on the qubit of Alice $q_a$ (or Bob $q_b$), where the gate $\mathcal{V}$ acts on.
        Then, the state can be realized by 
        \begin{align}
            \rho_{\mathrm{out}} & = \sum_{i,j,k,l} \sum_{\alpha, \beta, \mu, \nu} M_{j\beta,i\alpha} M_{l\nu,k\mu} \nonumber \\
            &~~~~\times \rho(A|i_{\alpha},l_{\nu}) \otimes \rho(B|k_{\mu},j_{\beta}),
        \end{align}
        where the unnormalized states $\rho(A|i_{\alpha},l_{\nu})$, $\rho(B|k_{\mu},j_{\beta})$ are
        \begin{align}
            \rho(A|i_{\alpha},l_{\nu}) & = \mathrm{Tr}_{q_a} [\hat{\Pi}_{i\alpha}^{(q_a)} \mathcal{V}_{2A}^{(q_c\rightarrow q_a)} \nonumber \\
            &~~~~\circ\mathcal{V}_{1A}^{(A)}(\rho_{A} \otimes \hat{\Pi}_{l\nu}^{(q_c)})], \\
            \rho(B|k_{\mu},j_{\beta}) & = \mathrm{Tr}_{q_d} [\hat{\Pi}_{k\mu}^{(q_d)}\mathcal{V}_{2B}^{(B)} \circ \mathcal{V}^{(q_d \rightarrow q_a)} \nonumber \\
            &~~~~\circ \mathcal{V}_{1B}^{(B)}(\rho_{B} \otimes \hat{\Pi}_{j\beta}^{(q_d)})], 
        \end{align}
        and the transition matrix is 
        \begin{equation}
            M_{j\beta,i\alpha} = (-1)^{\delta_{\alpha,y}} \delta_{\alpha,\beta} [\delta_{i,j}(1-\delta_{\alpha,y}) + (1-\delta_{i,j})\delta_{\alpha,y}].
        \end{equation}
        Here, $q_c$ and $q_d$ are two auxiliary qubits, and the notation $[\cdot]^{(q_{c,d} \rightarrow q_a)}$ represents that the operation $[\cdot]$ is acted by exchanging $q_a$ and $q_{c,d}$
        \begin{equation}
            [\cdot]^{(q_{c,d} \rightarrow q_a)} = [\cdot] \circ \mathcal{S}_{q_{c,d} \leftrightarrow q_a}.
        \end{equation}

        This method splits the nonlocal circuit into two separate circuits with two auxiliary qubits.
        The diagram is shown in Fig.~\ref{fig: division}. 
        The auxiliary qubits $q_c, q_d$ are prepared randomly in $\ket{i}_{\alpha}$, and the qubits $q_a$ and $q_d$ are measured randomly in $x,y$ and $z$ direction.
        After the preparations, evolutions, and measurements in $q_a$ and $q_d$, they obtain the states $\rho(A|i_{\alpha},l_{\nu})$ and $\rho(B|k_{\mu},j_{\beta})$, which are labeled by the prepared states $\ket{i_{\alpha}}, \ket{k_{\mu}}$, and the outcome $\ket{j_{\beta}}, \ket{l_{\nu}}$ of measurements.
        In the post-processing, they pairs the state $\rho(A|i_{\alpha},l_{\nu})$ with $\rho(B|k_{\mu},j_{\beta})$ and signature $\epsilon_{\alpha} \epsilon_{\nu}$.
        The index $\ket{i_{\alpha}}$ is related to $\bra{j_{\beta}}$, and $\ket{k_{\mu}}$ is related to $\ket{l_{\nu}}$, in the way that for a pair $(\ket{i_{\alpha}}, \bra{j_{\beta}})$, $i = j$ if $\alpha = \beta = x,z$ and $i \neq  j$ if $\alpha = \beta = y$.
        The signature $\epsilon_{\alpha} = (-1)^{\delta_{\alpha,y}}$ is negative when $\alpha = y$.
        This construction is from the equivalence between the teleportation forged by separable states and the standard QST shown in the previous.

        For a circuit with more nonlocal gates between Alice and Bob, the circuit can be split into separated circuits with the same method.
        The auxiliary qubits are $2 L$~for splitting $L$~nonlocal gates.
        Moreover, if the qubit $q_a$ can be reset, the qubit $q_c$ can reuse $q_a$ after the measurement, thus only $L$~auxiliary qubits are necessary.
        However, in compensation, the overhead is $3^{2L}$~times, which is optimal since the preparation in state $\ket{i_{\alpha}}$ (and $\ket{k_{\mu}}$) and measurement in $\bra{l_{\nu}}$ (and $\bra{j_{\beta}}$) on Alice's (and Bob's) side have no quantum correlation.

    \section{Mitigation of the Noise in Measurements} \label{sec: noise}
    
        The perfect realizations of both the methods discussed in the previous section depend on perfect preparations of state and measurements.
        In practice, there always exist errors in preparations and measurements.
        These errors can be canceled by the measurement error mitigation~\cite{RevModPhys.95.045005,PhysRevA.103.042605}.
        With the post-processing of measurement error mitigation, the overhead will increase additionally.
        In the following, we consider the cancellation of measurement errors in detail.

        Assume the expectation of observables of interest can be calculated from the projective measurement in basis $\ket{s_A} \otimes \ket{s_B}$, where $s_A$ and $s_B$ are bit strings.
        Then, the probabilities are 
        \begin{align}
            P(s_A, s_B) & = \sum_{i,j,k,l} \sum_{\alpha, \beta, \mu, \nu} M_{j\beta,i\alpha} M_{l\nu,k\mu} \nonumber \\
            &~~~~\times P(s_A|i_{\alpha},l_{\nu}) P(s_B|k_{\mu},j_{\beta}),
        \end{align}
        where 
        \begin{align}
            P(s_A|i_{\alpha},l_{\nu}) & = \frac{P(s_A, i_{\alpha}, l_{\nu})}{\sum_{s_A} P(s_A, i_{\alpha}, l_{\nu})}, \\
            P(s_B|k_{\mu},j_{\beta}) & = \frac{P(s_B, k_{\mu},j_{\beta})}{\sum_{s_B} P(s_B, k_{\mu},j_{\beta})},
        \end{align}
        where $P(s_A, i_{\alpha}, l_{\nu})$ and $P(s_B, k_{\mu},j_{\beta})$ can be counted from measurement data of experiments.

        In measurement error mitigation, there is an assignment matrix $\hat{A}$, which depicts the relation between the probabilities of ideal outcome and noisy outcome
        \begin{equation}
            P_{\mathrm{error}}(s) = \sum_{s'} A(s,s') P(s'),
        \end{equation}
        where $P$ and $P_{\mathrm{error}}$ are the ideal and noisy probabilities of outcome.
        This matrix can be measured from experiments.
        The error is mitigated by 
        \begin{equation}
            P(s) = \sum_{s'} A^{-1}(s,s') P_{\mathrm{error}}(s').
        \end{equation}
        In this manner, with the noisy probabilities $P_{\mathrm{error}}(s_A, i_{\alpha}, l_{\nu})$ and $P_{\mathrm{error}}(s_B, k_{\mu},j_{\beta})$ from experiments, the ideal probabilities is calculated by 
        \begin{widetext}
        \begin{align}
            P(s_A|i_{\alpha},l_{\nu}) & = \frac{\sum_{s_A', i_{\alpha}', l_{\nu}'} A^{-1}(s_A, i_{\alpha}, l_{\nu};s_A', i_{\alpha}', l_{\nu}') P_{\mathrm{error}}(s_A', i_{\alpha}', l_{\nu}')}{\sum_{s_A} \sum_{s_A', i_{\alpha}', l_{\nu}'} A^{-1}(s_A, i_{\alpha}, l_{\nu};s_A', i_{\alpha}', l_{\nu}') P_{\mathrm{error}}(s_A', i_{\alpha}', l_{\nu}')}, \\
            P(s_B|k_{\mu},j_{\beta}) & = \frac{\sum_{s_B', k_{\mu}',j_{\beta}'} A^{-1}(s_B, k_{\mu},j_{\beta};s_B', k_{\mu}',j_{\beta}') P_{\mathrm{error}}(s_B', k_{\mu}',j_{\beta}')}{\sum_{s_B} \sum_{s_B', k_{\mu}',j_{\beta}'} A^{-1}(s_B, k_{\mu},j_{\beta};s_B', k_{\mu}',j_{\beta}') P_{\mathrm{error}}(s_B', k_{\mu}',j_{\beta}')}.
        \end{align} 
        \end{widetext}
    
    \section{Conclusion} \label{sec: conclusion}

        In this paper, we demonstrate two methods for distributed quantum computation on separated quantum processors with local operations and classical channels.
        There is a simple method that is based on the teleportation protocol with classical forged entanglement.
        It requires $4$~auxiliary qubits and $9$~times overhead for classically teleporting per nonlocal gate. 
        However, the overhead of this construction may not be optimal.

        Moreover, we construct another method that requires $2$~auxiliary qubits per gate, which is less than the first one.
        This method is based on the identical operations forged by projective measurements.
        The demonstrated construction requires also $9$~times overhead, however, which is optimal.
        We also show that the measurement error mitigation technique can cancel the error of the protocol in noisy cases.

        Our results demonstrate the methods to implement a nonlocal quantum circuit on two separate quantum processors only with local operations and classical channels, which extend the possibility of combining two quantum processors.
        We expect that our methods will complement the toolbox of distributed quantum computation, and facilitate the extension of the scale of quantum computations. 

        \begin{acknowledgments}
            This work was supported by the Innovation Program for Quantum Science and Technology (Grant No. 2021ZD0301800), National Natural Science Foundation of China (Grants Nos. T2121001, 92265207, 12122504). We also acknowlege the supported from the Synergetic Extreme Condition User Facility (SECUF) in Beijing, China.
        \end{acknowledgments}
    
%
    
\end{document}